\documentclass{article}

\usepackage{microtype}
\usepackage{graphicx}
\usepackage{subfigure}
\usepackage{booktabs} 
\usepackage{amssymb}
\usepackage{hyperref}

\usepackage[accepted]{icml2024}

\usepackage{amsmath}
\usepackage{amssymb}
\usepackage{mathtools}
\usepackage{amsthm}

\usepackage[capitalize,noabbrev]{cleveref}

\theoremstyle{plain}

\theoremstyle{definition}

\theoremstyle{remark}

\usepackage[textsize=tiny]{todonotes}

\icmltitlerunning{MPCPA: Multi-Center Privacy Computing with Predictions Aggregation based on DDPM}

\begin{document}

\twocolumn[
\icmltitle{MPCPA: Multi-Center Privacy Computing with Predictions  \\ 
 Aggregation based on Denoising Diffusion Probabilistic Model }
 



\begin{icmlauthorlist}
\icmlauthor{Guibo Luo}{yyy}
\icmlauthor{Hanwen Zhang}{yyy}
\icmlauthor{Xiuling Wang}{steven}
\icmlauthor{Mingzhi Chen}{yyy}
\icmlauthor{Yuesheng Zhu}{yyy}

\end{icmlauthorlist}

\icmlaffiliation{yyy}{School of Eletronic and Computer Engineering, Peking University, Shenzhen, China}
\icmlaffiliation{steven}{Deparment of Computer Science, Stevens Institute of Technology, New Jersy, USA}
\icmlcorrespondingauthor{Xiuling Wang}{xwang193@stevens.edu}
\icmlcorrespondingauthor{YueSheng Zhu}{zhuys@pku.edu.cn}


\vskip 0.3in
]



\printAffiliationsAndNotice{}  

\begin{abstract}
Privacy-preserving computing is crucial for multi-center machine learning in many applications such as healthcare and finance. In this paper a Multi-center Privacy Computing framework with Predictions Aggregation (MPCPA) based on denoising diffusion probabilistic model (DDPM) is proposed, in which conditional diffusion model training, DDPM data generation, a classifier, and strategy of prediction aggregation are included. Compared to federated learning, this framework necessitates fewer communications and leverages high-quality generated data to support robust privacy computing. Experimental validation across multiple datasets demonstrates that the proposed framework outperforms classic federated learning and approaches the performance of centralized learning with original data. Moreover, our approach demonstrates robust security, effectively addressing challenges such as image memorization and membership inference attacks. Our experiments underscore the efficacy of the proposed framework in the realm of privacy computing, with the code set to be released soon.
\end{abstract}

\section{Introduction}
\label{submission}

In the past decade, machine learning, particularly artificial intelligence methods represented by deep learning, has been successfully applied to various tasks including but not limited to classification, recognition, segmentation, and generation \cite{schmidhuber_2015, LeCun_Bengio_Hinton_2015}. However, deep learning models typically require large-scale, diverse samples to train high-performance models. Collecting large labeled datasets is usually both challenging and costly. Additionally, in certain fields such as finance and healthcare, establishing a centralized multi-center dataset may encounter numerous restrictions including legal, privacy, technical, and data ownership concerns, especially requiring compliance with international laws and regulations (such as the European Union’s GDPR \cite{Houser_Voss_2018}, the United States’ HIPAA \cite{Pewen_2022}).

One method to overcome these issues is through federated learning \cite{konečný_mcmahan_yu_richtárik_suresh_bacon_2016,yang_liu_chen_tong_2019}, a training approach that enables distributed model training among multiple clients possessing local data, ensuring both data privacy protection and effective model training without exchanging local sample data. Instead, communication occurs solely through the exchange of model parameters or intermediate results, with Federated Averaging (Fed-Avg) \cite{mcmahan_moore_ramage_hampson_arcas_2016} being a typical method in this regard.

Classical federated learning requires multiple clients to engage in multi-round parameter or gradient iterations, leading to increased communication rounds becoming a bottleneck for platform setup and training speed \cite{mothukuri_parizi_pouriyeh_huang_dehghantanha_srivastava_2021}. Furthermore, Fed-Avg assumes that the data distribution between client clients is approximately independent and identically distributed (IID) \cite{zhao_li_lai_suda_civin_chandra_2018,zhang_wang_zhou_wu_zhang_2021} to achieve convergence during training and reach performance comparable to centralized training. However, when dealing with real-world Non-IID datasets, model performance might deteriorate, and the training might fail to converge.

In recent years, there has been significant progress in the development of neural network-based generative models. Generative models such as Variational Autoencoders (VAE) \cite{kingma_welling_2013}, Generative Adversarial Network (GAN) \cite{goodfellow_pouget-abadie_mirza_xu_warde-farley_ozair_courville_bengio_2017}, Flow model \cite{rezende_mohamed_wierstra_2014}, and Denoising Diffusion Probabilistic Model (DDPM) \cite{ho_jain_abbeel_2020} since 2021, have been widely employed in various data generation tasks. Several studies \cite{bindschaedler_shokri_gunter_2017,takagi_takahashi_cao_yoshikawa_2020} have indicated that data generated from these models holds the potential to be employed for safeguarding data privacy.

Probabilistic diffusion models have demonstrated the capacity to generate high-quality data \cite{Kazerouni_Aghdam_Heidari_Azad_Fayyaz_Hacihaliloglu_Merhof_2022} \cite{dhariwal_nichol_2021}. The application of DDPM data generation methods has recently been utilized to privacy-preserving computations \cite{yang_su_li_xue_2023,zhang_qi_zhao_2023} in multi-center datasets. However, these methods typically rely on pre-trained models, limiting their practical applicability to natural images. They often require additional techniques such as fine-tuning or distillation learning to enhance performance, resulting in insufficient versatility. Moreover, while DDPM is used to generate realistic data to complement the original federated learning framework, it does not fundamentally alter the underlying privacy computation framework. 

To establish a more versatile privacy computation framework, privacy computations based on DDPM-generated data must address the following core challenges:

(1) Training stable models with a small number of samples is a crucial challenge for DDPM, especially when starting from scratch without pre-trained models, aiming to generate stable and high-quality models for different datasets.

(2) Although diffusion-generated samples exhibit high quality, training a model solely on generated data to achieve performance similar to models trained on original data remains challenging.

(3) Whether the framework can achieve high-performance privacy computation with only a small number of communication rounds.

Moreover, validation is necessary to assess the model's generalization under Non-IID datasets, and its effectiveness in withstanding image memorization and membership inference attacks within the privacy computation framework.

To address these challenges, this paper proposes a general multi-center privacy computation framework based on DDPM data generation and ensemble learning. The framework includes conditional diffusion model training, DDPM data generation, a classifier, and predictions aggregation. Initially, each client trains a conditional diffusion model based on its own dataset and shares the DDPM model with other clients. Subsequently, each client trains a classification model using both its local original data and data generated by DDPM models from other clients. Finally, the central server aggregates the predictions of all clients’ models by ensemble learning.

The main contributions of this article are as follows:\\
(1) \textbf{Multi-center privacy computation framework.} The proposed multi-center privacy computation framework provides a general solution for privacy computation applicable to various types of datasets. Our framework effectively addresses the Non-IID problem encountered in federated learning while significantly reducing communication costs. Moreover, it demonstrates robust security, capable of resisting image memorization and membership inference attacks.

(2) \textbf{Classical and concise application of DDPM.} The utilization of DDPM generation models is both classical and concise in this paper. It does not require pre-trained models or complex parameter adjustments. Compared to GANs, the training process is simpler, making this framework suitable for generating high-quality data across various data types.

(3) \textbf{Maximizing data utilization while preserving privacy.} Each client maximizes the utilization of its original data and data generated by DDPM models from other clients, ensuring maximum utilization of local data while maintaining model performance stability without compromising privacy.

(4) \textbf{Performance enhancement by ensemble learning.} Ensemble learning of multiple clients’ predictions further enhances performance and sometimes even surpasses models trained on original data. 

\section{Related Work}
\subsection{Federated Learning}
Federated learning gained attention in 2015 \cite{konečný_mcmahan_ramage_2015} and has since evolved into a preferred privacy computing framework. However, the training process in federated learning encounters challenges such as high communication costs and issues related to Non-IID datasets.

The simplest and most direct solution to address communication overhead is to sacrifice model accuracy by training only low-capacity models that occupy a smaller communication space \cite{hamer_mohri_suresh_2020, caldas_konečný_mcmahan_talwalkar_2018}. FedBoost \cite{hamer_mohri_suresh_2020} primarily achieves federated ensembles by combining pre-trained base predictors with high performance and training efficiency.

One-shot Federated Learning (OSFL) has garnered widespread attention in recent years due to its lower communication costs. Most existing OSFL methods require the use of auxiliary datasets. \cite{guha_talwalkar_smith_2019} utilized unlabeled public data on the server for model expansion. Lin et al. \cite{Lin_Kong_Stich_Jaggi_2020} suggested leveraging auxiliary datasets for knowledge transfer on the server. With significant advancements in deep generative models, recent work have explored generative OSFL methods. DENSE \cite{zhang_chen_li_lyu_wu_ding_shen_wu_2021} employed a collection of client models as a discriminator to train a generator for generating pseudo-samples, which are then used to train the aggregated model. FedCVAE \cite{heinbaugh2022data} trained conditional variational autoencoders (CVAE) on client-side, sending the decoder to the server for generating global data used in training the classifier. FedCVAE effectively integrates information from different client sources, reducing communication costs, and preserving data privacy. However, the limited generative capacity of CVAE may potentially restrict the model's performance on more complex datasets. CVAE focuses on using entirely synthetic samples to train models, neglecting the potential of private real data from individual clients in training. FedCADO \cite{yang_su_li_xue_2023} utilized a classifier-guided diffusion model to generate data aligned with client distributions, followed by training an aggregated model on the server within the one-shot federated learning framework.

Traditional federated learning methods exhibit low efficiency when dealing with heterogeneous data \cite{zhao_li_lai_suda_civin_chandra_2018}, and the performance of models may degrade when handling real-world Non-IID datasets \cite{hsu_qi_brown_2019,karimireddy_kale_mohri_reddi_stich_suresh_2019}. In some cases, training may even fail to converge. Several recent works have been dedicated to addressing the issue of non-IID data in federated learning, including FedProx \cite{tian_sahu_zaheer_sanjabi_talwalkar_smith_2018}, SCAFFOLD \cite{karimireddy_kale_mohri_reddi_stich_suresh_2019}, and FedNova \cite{wang_liu_liang_joshi_poor_2020}. Data-based methods preprocess client data before training \cite{shin_hwang_kim_park_bennis_kim_2020}, making the distribution of client data as similar as possible to the overall data distribution. Personalized federated learning can optimize the local model for different users \cite{Kulkarni_Kulkarni_Pant_2020}. Overall, Non-IID data poses significant challenges to the learning accuracy of federated learning algorithms, and there is no one-size-fits-all advanced federated learning algorithm that outperforms others in all aspects \cite{li_diao_chen_he_2022}.

\subsection{Learning From Synthetic Data}

Recently, there has been a rise in privacy computing initiatives leveraging deep generative models, employing them to generate synthetic training data as a solution to privacy concerns \cite{bindschaedler_shokri_gunter_2017}\cite{takagi_takahashi_cao_yoshikawa_2020}. \cite{wei_kreavcic_wang_yin_chien_potluru_li_2023} mathematically described the privacy guarantee of data generated by a diffusion model in the context of synthetic data sharing. In 2020, AsynDGAN \cite{chang_qu_zhang_sabuncu_chen_zhang_metaxas_2020} used a distributed asynchronous discriminator to learn the distribution of real images without sharing raw patient data from different datasets. In 2023, the Distributed Synthesis Learning (DSL) \cite{chang_yan_zhou_qu_zhang_baskaran_al’aref_zhang_metaxas_2022} architecture was proposed to learn and use synthetic images across multiple medical centers, ensuring the protection of sensitive personal information. However, both AsynDGAN and DSL architecture requires frequent communication during the whole training process. In each iteration, DSL transfers synthesized images, masks, and gradients between the central generator and medical entities, incurring significant communication costs. Furthermore, GANs fall behind Diffusion models in terms of sample generation quality and training stability, leaving room for improvement.

\begin{figure*}[ht]
\vskip 0.2in
\begin{center}
\centerline{\includegraphics[width=0.9\linewidth]{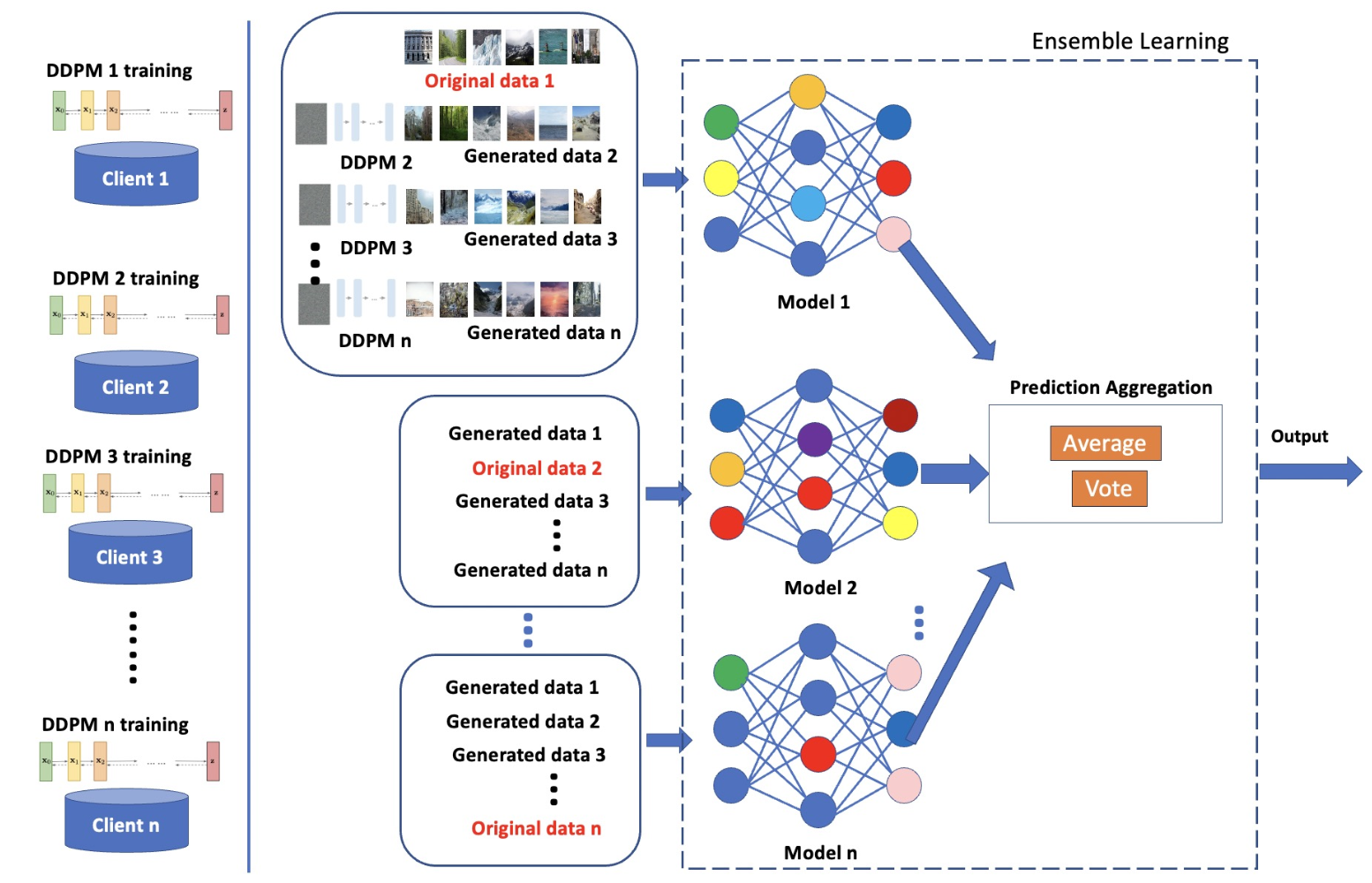}}
\caption{The proposed multi-center privacy computing framework, which consists of conditional diffusion model training, DDPM data generation, classifiers, and ensemble learning modules}
\label{fig-1}
\end{center}
\vskip -0.2in
\end{figure*}

\section{Method}

Our goal is to establish a universal, generalized multi-center asynchronous privacy computation framework. The proposed framework serves as a universal privacy computation solution, which includes conditional diffusion model training, DDPM data generation, classifiers, and ensemble learning modules as shown in \cref{fig-1}. 

Most current image generation efforts are focused on using pre-trained diffusion models to generate images, which often rely on large-scale publicly available datasets. In contrast, our DDPM module employs a classical and concise conditional DDPM model. This allows for the training of stable and high-quality generative models from scratch, tailored to the unique datasets of various clients. It demonstrates robust generative performance, even when dealing with small sample sizes. This approach is well-suited to accommodate the diverse data types of each client, not limited to natural images, making it more versatile and easier to generalize. 

Once each client completes its training, transmitting the conditional diffusion model to other clients through a central server. 

Federated learning faces challenges in practical applications where multiple clients are involved in multiple rounds of parameter or gradient iteration, thereby constraining platform setup and training speed. To address these limitations, we utilize an ensemble learning module for predictive aggregation, eliminating the need for multiple rounds of iteration.

\begin{figure*}[ht]
\vskip 0.2in
\begin{center}
\centerline{\includegraphics[width=0.9\linewidth]{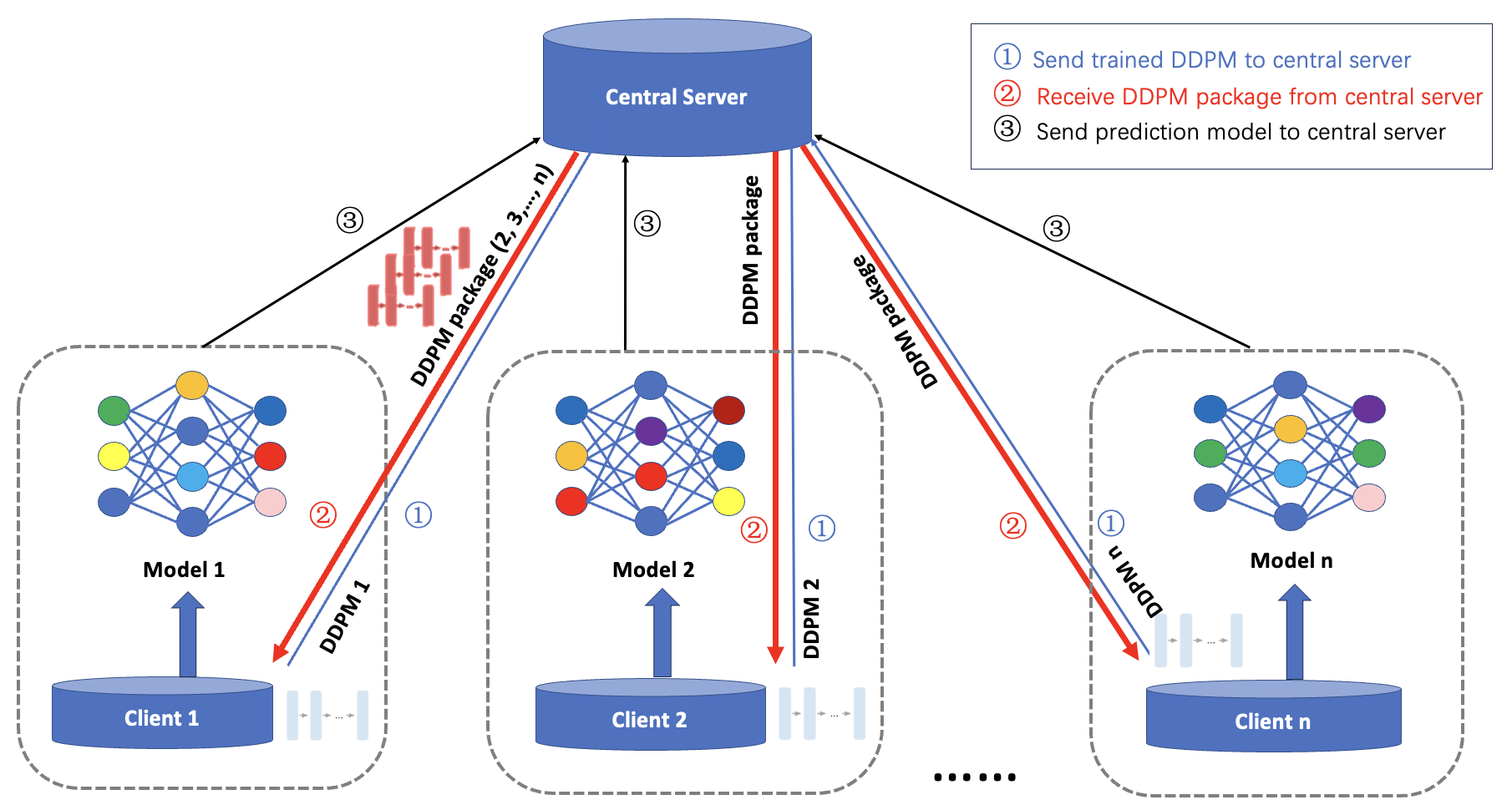}}
\caption{Communications between clients and central server.}
\label{fig-2}
\end{center}
\vskip -0.2in
\end{figure*}

\subsection{DDPM Generation Module}

The objective of conditional generative models is to generate a target image $x_0$ given a condition $y$. The diffusion model comprises two processes: the forward noisy process $q$ and the reverse denoising process $p_\theta$. In the forward noisy process, the image $x_0$ from the training set undergoes $T$ rounds of noise addition to make $x_T$ comply with a standard normal distribution. The noise addition process $q$ progressively alters the original image from slow to fast. This process is a Markovian noising process, where Gaussian noise is added to the image $x_{t-1}$ at each time step $t=1,2, \ldots, T$ according to a variance schedule $\beta_t:$
$$
\mathbf{x}_t=\sqrt{1-\beta_t} \mathbf{x}_{t-1}+\sqrt{\beta_t} \mathcal{N}(\mathbf{0}, \mathbf{I})\eqno{(1)}
$$
where $N(\cdot)$ denotes the normal distribution and $\mathbf{I}$ is the identity matrix. Equation (1) can also be expressed in the form of a probability distribution:
$$
q\left(\mathbf{x}_t \mid \mathbf{x}_{t-1}\right)=\mathcal{N}\left(\mathbf{x}_t ; \boldsymbol{\mu}_t=\sqrt{1-\beta_t} \mathbf{x}_{t-1}, \mathbf{\Sigma}_t=\beta_t \mathbf{I}\right)\eqno{(2)}
$$
In the reverse process, we aim to reverse each step of adding noise, restoring a noisy image back to an image from the dataset. Given the condition $y$, the reverse process learns to denoise the sample $x_t$ by one step to $x_{t-1}$. This can be represented as
$$
q\left(x_{t-1} \mid x_t\right)=q\left(x_t \mid x_{t-1}\right) \frac{q\left(x_{t-1}\right)}{q\left(x_t\right)}\eqno{(3)}
$$
The reverse process $p_\theta$ is difficult to solve theoretically but can be approximated parameterized as
$$
p_\theta\left(\mathbf{x}_{t-1} \mid \mathbf{x}_t\right)=\mathcal{N}\left(\mathbf{x}_{t-1} ; \mu_\theta\left(\mathbf{x}_t, t\right), \Sigma_\theta\left(\mathbf{x}_t, t\right)\right)\eqno{(4)}
$$

Each step of the noise-reversing operation follows a normal distribution, and when given a certain input, the mean and variance of this normal distribution can be expressed analytically with a mean of $\mu_\theta\left(\mathbf{x}_t, t\right)$ and a variance of $\Sigma_\theta\left(\mathbf{x}_t, t\right)$. The learning objective of the neural network is to make its output distribution consistent with the theoretically calculated distribution. In this way, starting from the Gaussian noise $x_T \sim N(0,1)$, and given $\mathrm{y}$, we can iteratively infer the sample at time step $t-1$ from the sample at time step $t$ until we reach the original image $x_0$
Through some mathematical simplifications, the problem is transformed into fitting the random noise $\epsilon_t$ at time step $\mathrm{t}$ used in generating $\mathbf{x}_t$. The training objective can be rewritten as
$$
\begin{gathered}
L_t^{\text {simple }}=\mathbb{E}_{t \sim[1, T], \mathbf{x}_0, \epsilon_t}\left[\left\|\epsilon_t-\epsilon_\theta\left(\mathbf{x}_t, t\right)\right\|^2\right]\\
=\mathbb{E}_{t \sim[1, T], \mathbf{x}_0, \epsilon_t}\left[\left\|\epsilon_t-\epsilon_\theta\left(\sqrt{\bar{\alpha}_t} \mathbf{x}_0+\sqrt{1-\bar{\alpha}_t} \epsilon_t,t\right)\right\|^2\right]
\end{gathered}
\eqno{(5)}
$$
where $\epsilon_\theta\left(\boldsymbol{x}_t, y, t\right)$ is a function approximating $\epsilon_t$.

In DDPM training, any neural network architecture can be used to fit the reverse denoising process theoretically. However, due to the similarity between DDPM tasks and image denoising tasks, DDPM has opted for U-Net \cite{ronneberger_fischer_brox_2015} as the model structure for noise prediction. U-Net is a U-shaped network structure composed of an encoder, a decoder, and skip connections between the encoder and decoder. The encoder downsamples the image into a feature representation, while the decoder upsamples this feature into the target noise, using cross-layer connections to concatenate the features between the encoder and decoder.

\subsection{Prediction Aggregation}

During the predictive aggregation phase, the ensemble learning module is used to fuse classifiers trained by each client using their local raw data and other clients' synthesized data, thereby enhancing the performance of the classification task. Ensemble learning accomplishes learning tasks by constructing and combining multiple learners. One of the main reasons for the success of ensemble methods is the diversity of base learners \cite{dietterich_2000}. In multi-center learning, due to the influence of data heterogeneity, different clients use different datasets to train base classifiers, and the errors of the trained base classifiers can be considered independent of each other. This allows us to achieve more stable and reliable results through the fusion of strategies. The theoretical foundation of ensemble learning is the multi-dataset bias-variance-covariance decomposition \cite{dietterich_2000}, which breaks down the generalization error of learners into three parts: bias, variance, and covariance. The details of multi-dataset bias-variance-covariance decomposition can be found in Appendix\ref{app:A.1}.

From the perspective of aggregating base learners, ensemble learning can be broadly categorized into the following two strategies.

Averaging Method: The averaging method is the most common approach in ensemble learning, which can be divided into simple averaging and weighted averaging. For the simple averaging method, the ensemble model aggregates individual base learners with equal weights to obtain the final prediction: $${Y(x)=\frac{1}{n} \sum_1^n y_i(x)}\eqno{(6)}$$
As deep learning architectures exhibit high variance and low bias, the simple averaging of the ensemble model improves generalization performance due to the reduction in variance between models \cite{ganaie_hu_malik_tanveer_suganthan_2022}. Apart from simple averaging, different weighting methods can be used to determine the weights of the base learners for weighted averaging: determining weights based on the multivariate Gaussian distribution to calculate weights from near to far; using the Delphi method based on expert ratings; weights can be calculated from the consistency matrix through the Analytic Hierarchy Process (AHP); estimation of detection and state transition in the Kalman filter can also calculate weights.

Voting Method: Similar to simple averaging, the voting method combines the outputs of the base learners. However, instead of taking the average of the predictions, the voting method calculates the votes of the base learners and predicts the final label as the label with the majority of votes. The voting method follows the principle of majority rule in ensemble learning, reducing variance by aggregating multiple models to enhance model robustness. The voting method can be categorized based on the calculation of votes into absolute majority voting, relative majority voting, and weighted voting.

\subsection{Communications of MPCPA}

This framework only requires three stages of communication between the clients and server as described in \cref{fig-2}: transmitting the DDPM model and transmitting the prediction model. In the phase of transmitting the DDPM model, each client needs to go through the upload and download of the DDPM model from central server. During the prediction model transmission phase, each client needs to transmit the prediction model to the central server. The detailed communication process is illustrated in the following.

(1) Upload of DDPM models:
Each client sends the trained conditional diffusion model from its local private dataset to the central server. Assuming there are $n$ clients, it requires a total of $n$ transmissions.

(2) Download of DDPM package:
Each client downloads the collection of DDPM models from the central server, referred to as the DDPM package. Assuming there are $n$ clients, it requires a total of $n$ transmissions. 

(3) Transmission of prediction models:
During the prediction aggregation phase, all prediction models are transmitted to the server, which requires a total of $n$ transmissions.

Therefore, considering the scenario with a central server, the total communication times does not exceed $3n$ times.

\begin{algorithm}[tb]
   \caption{MPCPA}
   \label{alg:example}
\begin{algorithmic}
   \STATE {\bfseries procedure} \MakeUppercase{client}
   \FOR{each client $k\in\ I$}
   \STATE {$f_k(y) \leftarrow \boldsymbol{D D P M T r a i n}\left(R_k, T_L, y\right)$}
   \STATE {upload $f_k(y)$ to server and download $P^k (y)$}
   \STATE {$P^k (y)={\{f_k(y)\}}_{i=1,2,\ldots,k-1,k+1,\ldots,n}$}
   \STATE {generate synthetic samples using $P^k(y)$\\
   $S^k = S_1\bigcup\ S_2\ldots\bigcup\ S_{k-1}\bigcup\ S_{k+1}\ldots\bigcup\ S_n$ \\}
   \STATE {combine local dataset and generated data 
   $D_k = S^k\bigcup\ R_k$ \\}
   \STATE {train classifier $C_k$}
   \FOR{classifier epoch from $t=1$ to $T_c$}
   \FOR{mini-batch $b\subset\ D_k$}
   \STATE {$\mathbf{w}_C^k\gets\mathbf{w}_C^k-\eta_C\cdot\nabla_{\mathbf{w}_C^k}\ell_C\left(\mathbf{w}_C^k;b\right)$}
   \ENDFOR
   \ENDFOR\\
   \STATE {upload classifier $C_k$ to the server $k$}
   \ENDFOR
   \STATE {\bfseries procedure} \MakeUppercase{server} \\
   \STATE {receive clients’ DDPM models ${\{f_k(y)\}}_{i=1,2,\ldots,k,\ldots,n}$}
   \FOR{client $k\in\ C$}
   \STATE {send $P^k (y)$ to client $k$}
   \ENDFOR\\
   \STATE {receive clients’ classifier models $C_1,C_2,\ldots,\ C_n$\\}
   \STATE {ensemble predict results of $n$ classifiers}
   \STATE {$\mathrm{y}_{\mathrm{E}}=\operatorname{Ensemble}\left(y_{C 1}, y_{C 2}, \ldots, y_{C n}\right)$}
\end{algorithmic}
\end{algorithm}

\subsection{Algorithms}
$I$ is the set of clients with $n$ clients in total. The local diffusion model training parameters are condition $y$, with training epochs $T_L$, local dataset $R_k$. $f_k (y)$ represents the trained local diffusion model and $P^k (y)$ is a package contains diffusion models from other clients. $S^k$  is the synthetic samples of client $k$ generated by $P^k (y)$, $D_k$  is the dataset for classifier training, which includes generated dataset and private local dataset. The classifier parameters are $w_C^k$, with training epochs $T_c$, classification loss $l_C(·)$, and learning rate $\eta_C$. $y_E$ represents ensemble prediction that aggregates by server. \cref{alg:example} shows how MPCPA works.


\section{Experiments}

\begin{table*}[t]
\caption{Accuracy (\%) of classification models trained with different training set and training method across IntelImage, NeoJaundice, ChestX-Ray Pneumonia.}
\label{table-1}
\vskip 0.15in
\begin{center}
\begin{small}
\setlength{\tabcolsep}{4mm}{\begin{tabular}{llcccccc}
\toprule
Training method & Dataset & \multicolumn{2}{c}{IntelImage}&  \multicolumn{2}{c}{NeoJaundice} &\multicolumn{2}{c}{ChestX-Ray}\\
~ & Training set & validation & test & validation & test & validation & test \\
\midrule
 Centralized learning & All original data & 91.63 & 92.98 & 79.66 & 76.30 & 90.49 & \textbf{91.28} \\
Centralized learning & All generated data & 87.25 & 88.12 & 78.22 & 76.30 & 90.14 & 90.69 \\
Federated learning & original data & 91.24 & 92.44 & 76.79 & 77.03 & 86.06 & 88.37 \\
Ours (MPCPA) & original data \\~ & + generated data & \textbf{91.96} & \textbf{93.25} & \textbf{81.38} & \textbf{77.04} & \textbf{91.37} & 90.70 \\
\bottomrule
\end{tabular}
}
\end{small}
\end{center}
\vskip -0.1in
\end{table*}

\subsection{Dataset}
To validate the generalizability of our method across diverse image types, we conduct experiments on both natural and medical images, specifically utilizing the IntelImage, NeoJaundice, ChestX-Ray Pneumonia and Tuberculosis datasets. To assess performance in real-world Non-IID datasets, three Tuberculosis datasets are collected from various sites, including the Montgomery set, Shenzhen set, and India Set. Additionally, to evaluate performance on external dataset, another Tuberculosis dataset is served as an external set. Each dataset of the IntelImage, NeoJaundice, and ChestX-Ray Pneumonia is randomly partitioned into three clients. While it's possible to configure additional clients for increased numbers, for the sake of simplicity, we have chosen to use three clients in this setup. Training of both the DDIM models and classification tasks are conducted on the training data. Subsequently, the performance and results are assessed based on the validation and testing datasets. More details can be found in the Appendix\ref{app:B.1}.

\begin{table*}[t]
\caption{Accuracy (\%) of classification models trained with different combinations of training methods and training sets on Tuberculosis dataset.}
\label{table-2}
\vskip 0.15in
\begin{center}
\begin{small}
\begin{rm}
\setlength{\tabcolsep}{5mm}{\begin{tabular}{llcccccc}
\toprule
~ & Training set & validation & test & external set \\
\midrule
Centralized learning & All original data & 90.59 & 88.60 & 62.86 \\
& All generated data & \textbf{91.76} & 87.04 & 61.96 \\ 
& Shenzhen Set (A1) & 83.53 & 79.27 & 60.14 \\ 
& Shenzhen Set+other generated data (B1) & 89.41 & 85.49 & 62.50 \\
& IndiaSet (A2) & 56.47 & 54.41 & 39.67 \\
& IndiaSet+other generated data (B2) & 88.24 & 84.05 & 56.88 \\
& MontgomerSet (A3) & 68.24 & 71.50 & 53.80 \\
& MontgomerSet+other generated data (B3) & 88.24 & 83.94 & 65.76 \\
\addlinespace[0.5ex]
\hline
\addlinespace[0.5ex]
Federated learning & Individual’s original data & 85.88 & 83.42 & 59.24 \\ 
\hline
\addlinespace[0.5ex]
Predictions aggregation & (A1 to A3)  & 75.29 & 80.31 & 60.69 \\
& (B1 to B3) (Ours) & 90.59 & \textbf{90.16} & \textbf{63.77} \\
\bottomrule
\end{tabular}
}
\end{rm}
\end{small}
\end{center}
\vskip -0.1in
\end{table*}

\subsection{Classification Results}
The compared methods include centralized learning and federated learning. DDPM-based one-shot federated learning is excluded due to its reliance on diffusion pre-trained models, currently available only for natural images, with no corresponding pre-trained models for medical data. Our overarching objective is to validate the effectiveness of our privacy computation framework and confirm its performance comparability to centralized learning. The experiment details can be found in the Appendix\ref{app:B.2}.

\cref{table-1} demonstrates different combinations of training methods (centralized learning, our proposed method - MPCPA, and federated learning) and training sets (original data and generated data) across IntelImage, NeoJaundice, and ChestX-Ray Pneumonia datasets. In the IntelImage dataset, our proposed MPCPA method exhibits a slight improvement over both centralized learning and federated learning using original data. In the NeoJaundice dataset, our method achieves superior overall performance, surpassing federated learning by approximately 5\% in the validation set. For ChestX-Ray Pneumonia, our approach demonstrates comparable accuracy to centralized learning with all original data, outperforming federated learning by approximately 5\% in the validation set and 2\% in the test set. Notably, models trained on generated data using centralized learning experience more significant performance loss compared to their all original data counterparts. Overall, our method consistently achieves results similar to or even outperform centralized learning with original data, and may be much better than federated learning in specific cases.

 \subsection{Non-IID Dataset Evaluation and External Testing}
 Addressing the challenge of Non-IID datasets is crucial for the Multi-center Privacy Computing method, as their presence can potentially degrade performance. In this section, we tackle this issue by selecting a real-world Non-IID dataset focused on Tuberculosis. The objective is to assess the impact of different combinations of learning methods and training sets on the model's performance. Additionally, external testing is employed to offer a more robust evaluation, analyzing how well the models generalize to previously unseen data from the respective training sets.
 
\cref{table-2} shows the performance of diverse training methods, including centralized learning, predictions aggregation, and federated learning, across various training sets on the Tuberculosis dataset. Models exclusively trained on single-client data (A1, A2, A3) exhibit suboptimal results, emphasizing the challenge posed by limited data size, particularly in the critical domain of medical imaging where robust generalization is essential. The combination of original data with generated data from other clients (B1, B2, B3) demonstrates a substantial improvement, exceeding 10\% on average. 

Despite observing some performance loss in models trained on all generated data or single site data plus generated data compared to using all original data, predictions aggregation across all clients' predictions consistently outperforms centralized learning with original data. Our approach outperforms federated learning by approximately 5\% on average across validation, test, and external sets. This improvement is attributed to the distinct distributions among the three Tuberculosis clients. Handling Non-IID data across different clients poses challenges for federated learning. However, our method's generated data effectively mitigates distribution differences among clients. Additional illustrative results can be found in the Appendix\ref{app:C.1}.

\begin{table}[t]
\caption{Communication times for federated learning and MPCPA.}
\label{table-3}
\vskip 0.15in
\begin{center}
\begin{small}
\setlength{\tabcolsep}{6.5mm}{\begin{tabular}{lc}
\toprule
Method & Communication times \\~&($n = 3, iters = 200$)\\
\midrule
federated learning & $2n * iters$ = 1200 \\
 MPCPA & $3n$ = 9 \\
\bottomrule
\end{tabular}
}
\end{small}
\end{center}
\vskip -0.1in
\end{table}

\subsection{Communications}
Federated learning requires a total of $2n\ \ast\ iters$ communications, whereas the proposed MPCPA requires only $3n$ communications. In our experiments, where $n=3$ and $iters= 200$, the communication times are presented in \cref{table-3}. It is evident that MPCPA significantly reduces the number of communications compared to federated learning. This reduction stems from the fact that there is no need for communication during each client task's training.

\subsection{Privacy Analysis}
\textbf{Image Memorization}

Diffusion models may memorize individual images from their training data and replicate specific images from their training dataset during generation \cite{Carlini_Hayes_Nasr_Jagielski_Sehwag_Tramer_Balle_Ippolito_Wallace_2023}, which raises concern about the privacy issue of diffusion model. Here, we adopt notion of memorization in \cite{balle_cherubin_hayes_2022}, defining the image memorization from the perspective of image distance function, and evaluating the extent of memorization on our diffusion model. $\ell$ is a distance function and $\delta$ is a threshold that determines whether two images as being identical. We say that an image $x$ in training set is memorized by a diffusion model $f_\theta$ if there exists an image $\hat{x}$ generated by $f_\theta$ has the property that $\ell(x, \hat{x}) \leq \delta$. For distance function, we use the Euclidean 2-norm distance $\ell_2(a, b)=\sqrt{\sum_i\left(a_i-b_i\right)^2 / d}$ where $d$ is the dimension of the inputs. 

A quantitative analysis is conducted on 3200 images from each class generated by each client's DDPM model. For every generated image, its minimum Euclidean 2-norm distance to the original training images is calculated. As shown in \cref{table-4}, the minimum distance for each DDPM model’s generated image are greater than the threshold $\delta$ of 0.1 suggested in \cite{Carlini_Hayes_Nasr_Jagielski_Sehwag_Tramer_Balle_Ippolito_Wallace_2023}. This finding substantiates that the transmission of DDPM model does not compromise the privacy of the client's private data.
\begin{table}[ht]
\caption{Image memorization measurement: The minimum distance for each DDPM model’s generated image to the original training images. (The abbreviations II, NJ, PM, and TB correspond to IntelImage, NeoJaundice, ChestX-Ray Pneumonia, and Tuberculosis, respectively.)}
\label{table-4}
\vskip 0.15in
\begin{center}
\begin{small}
\setlength{\tabcolsep}{3.2mm}{\begin{tabular}{lcccc}
\toprule
DDPM & II & NJ & PM & TB \\
\midrule
Client1 & 0.1761 & 0.4050 & 0.2888 & 0.3760 \\ 
Client2 & 0.2151 & 0.3837 & 0.2588 & 0.1165 \\ 
Client3 & 0.1576 & 0.3134 & 0.3088 & 0.2408 \\
\bottomrule
\end{tabular}
}
\end{small}
\end{center}
\vskip -0.1in
\end{table}

\textbf{Membership Inference Attacks}

The purpose of a membership inference attack (MIA) is to ascertain whether a given piece of data is part of the training set utilized in training a machine learning model. We evaluate membership inference with the loss threshold attack \cite{yeom_giacomelli_fredrikson_jha_2018}. The loss threshold attack technique is based on the fact that training examples have lower loss than non-training examples because models are trained to minimize their loss on the training set. The loss threshold attack thus computes the loss $l=\mathcal{L}(x ; f)$ and reports ``member'' if $l$ $<\tau$ for some chosen threshold $\tau$ and otherwise ``non-member''. 

\cref{table-5} shows the MIA accuracy for each classifier model. Notably, classifiers trained on both original and generated data demonstrate a similar MIA accuracy compared to the original and federated models. This observation indicates that our method does not exacerbate privacy leaks.
\begin{table}[ht]
\caption{Membership inference attacks accuracy (\%) for each classifier model.}
\label{table-5}
\vskip 0.15in
\begin{center}
\begin{small}
\begin{rm}
\setlength{\tabcolsep}{2.8mm}{\begin{tabular}{lcccc}
\toprule
Dataset & II & NJ & PM & TB \\
\midrule
Original data & 0.4428 & 0.6786 & 0.4786 & 0.6500 \\
All generated & 0.4857 & 0.5429 & 0.5357 & 0.5071 \\
A1 & 0.5714 & 0.6500 & 0.6000 & 0.5714 \\
B1 & 0.5857 & 0.6857 & 0.5000 & 0.5714 \\
A2 & 0.5786 & 0.7214 & 0.5860 & 0.5217 \\
B2 & 0.5357 & 0.6643 & 0.4714 & 0.5870 \\
A3 & 0.5643 & 0.7500 & 0.5714 & 0.6018 \\
B3 & 0.5857 & 0.6500 & 0.4857 & 0.6111 \\
Federated & 0.5500 & 0.8143 & 0.5429 & 0.6286 \\
\bottomrule
\end{tabular}
}
\end{rm}
\end{small}
\end{center}
\vskip 0.02in
A* denotes Client*; B* denotes ``Client* + other generated data''.
\end{table}

\textbf{More results can be found in Appendix\ref{app:C} and \ref{app:D}}.
\section{Conclusion}
This paper introduces a multi-center privacy computing framework that integrates DDPMs for data generation and predictions aggregation. The substitution of DDPMs for original data from other clients forms a key component for enhancing privacy preservation. Notably, this process significantly reduces the need for communication during each client task's training, differentiating it from classic federated learning. A comprehensive set of experiments is conducted across multiple datasets. The experimental outcomes demonstrate that the framework outperforms federated learning and approaches the accuracy achieved by centralized learning. Remarkably, when confronted with Non-IID datasets, the proposed framework exhibits a substantial improvement exceeding 5\% compared to federated learning. Predictions aggregation serves as a pivotal factor in enhancing overall performance, especially when combining original and generated data.

In conclusion, the proposed multi-center privacy computing framework, distinguished by DDPM-based data generation and predictions aggregation, proves to be a potent solution for privacy-preserving computing in decentralized settings.

\nocite{langley00}

\bibliography{icml_2024}
\bibliographystyle{icml2024}

\newpage
\appendix
\onecolumn
\section{Appendix}
\label{app:A}
\subsection{Theoretical foundation of ensemble learning}
\label{app:A.1}

The theoretical foundation of ensemble learning is the multi-dataset bias-variance-covariance decomposition \cite{dietterich_2000}, which breaks down the generalization error of learners into three parts: bias, variance, and covariance. Without loss of generality, assuming equal weights for each learner, the square error of the ensemble can be mathematically represented as:
$$\begin{aligned} E[o-t]^2 & =\text { bias }^2+\frac{1}{M} \operatorname{var}+\left(1-\frac{1}{M}\right) \text { covar }, \\ \text { bias } & =\frac{1}{M} \sum_i\left(E\left[o_i\right]-t\right), \\ \operatorname{var} & =\frac{1}{M} \sum_i E\left[o_i-E\left[o_i\right]\right]^2, \\ \text { covar } & =\frac{1}{M(M-1)} \sum_i \sum_{j \neq i} E\left[o_i-E\left[o_i\right]\right]\left[o_j-E\left[o_j\right]\right],\end{aligned}\eqno{(7)}$$
Bias measures the average estimation of how close the base learning algorithm can approximate the target. Variance measures the fluctuation of the estimated values of the base learning method for different training sets of the same size. In fact, the square error of the ensemble largely depends on the covariance term, which simulates the correlation between individual base learners. The smaller the covariance, the better the performance of the ensemble. Obviously, if all learners make similar errors, the covariance will be large, indicating that the diversity of base learners is crucial for the performance of ensemble learning.

\section{Experimental settings}
\label{app:B}
\subsection{Dataset}
\label{app:B.1}
\textbf{IntelImage}

IntelImage was originally released by Intel to conduct an Image Classification Challenge, featuring a dataset comprising approximately 25,000 images capturing natural scenes worldwide. The dataset consists of images with dimensions of 150x150 pixels, categorized into six classes: `buildings', `forest', `glacier', `mountain', `sea', and `street'. Specifically, the dataset is partitioned into approximately 14,000 images for training, 3,000 for validation, and 7,000 for testing purposes.

\textbf{NeoJaundice}

The NeoJaundice dataset \cite{Wang_Wang_Wang_Li_Da_Liu_Gao_Shen_He_Shen_etal._2023} is designed for neonatal jaundice research and is comprised of skin photos. It encompasses 2,235 images from 745 infants, with an average size of 567 × 567 pixels. Collected at Xuzhou Central Hospital, this dataset's binary labels were initially derived from total serum bilirubin readings sourced from the hospital's health information system, employing a threshold of 12.9 mg/dL. These labels were subsequently validated by a senior pediatrician with extensive experience. For each infant, three images were captured on different body skin areas—head, face, and chest—using digital cameras. To ensure color accuracy, the skin regions in the images were framed by a standardized color card, serving the purpose of color calibration.

\textbf{ChestX-Ray Pneumonia}

The ChestX-Ray Pneumonia dataset \cite{Kermany_Goldbaum_Cai_Valentim_Liang_Baxter_McKeown_Yang_Wu_Yan_etal._2018} is structured into three main folders (train, test, val) and further divided into subfolders for each image category, namely Pneumonia and Normal. It comprises a total of 5,863 chest X-ray images in JPEG format, categorized into two classes: Pneumonia and Normal. The images, captured in the anterior-posterior view, were collected from retrospective cohorts of pediatric patients aged one to five years at Guangzhou Women and Children’s Medical Center in Guangzhou. These chest X-ray images were part of the routine clinical care for patients. To ensure data quality, all chest radiographs underwent an initial screening process, eliminating low-quality or unreadable scans. The diagnoses for the remaining images were then assessed by two expert physicians before being used to train the AI system. To mitigate grading errors, a third expert independently reviewed the evaluation set.

\textbf{Tuberculosis}

To evaluate the performance in real world Non-IID dataset, three Tuberculosis dataset are collected from different sites. Additionally, to evaluate performance on external datasets, another Tuberculosis dataset are collected from an external source.

\textbf{Montgomery County chest X-ray set (MC)}

The MC dataset \cite{Candemir_Jaeger_Palaniappan_Musco_Singh_ZhiyunXue_Karargyris_Antani_Thoma_McDonald_2014} was collaboratively collected with the Department of Health and Human Services, Montgomery County, Maryland, USA. This dataset encompasses 138 frontal chest X-rays obtained from Montgomery County’s Tuberculosis screening program. Among these, 80 cases are categorized as normal, while 58 exhibit manifestations of Tuberculosis. The X-rays were captured using a Eureka stationary X-ray machine (CR) and are available in PNG format as 12-bit gray level images. The dimensions of the X-rays vary, with sizes either 4,020×4,892 or 4,892×4,020 pixels.

\textbf{Shenzhen chest X-ray set}

The Shenzhen dataset \cite{Jaeger_Candemir_Antani_Wáng_Lu_Thoma_2014} was collaboratively collected in partnership with Shenzhen No.3 People’s Hospital, Guangdong Medical College, Shenzhen, China. This dataset comprises 662 frontal chest X-rays, with 326 representing normal cases and 336 exhibiting manifestations of Tuberculosis, including pediatric X-rays in the anterior-posterior (AP) view. The X-rays are provided in PNG format, and while their size can vary, they are approximately 3,000 × 3,000 pixels.

\textbf{India chest X-ray set}

The Tuberculosis dataset from Jaypee University of Information Technology in India comprises 155 frontal chest X-rays. Among these, 77 cases are categorized as normal, while 78 exhibit manifestations of Tuberculosis. The X-rays are provided in JPG format, and while their size can vary, most images are 1024 × 1024 pixels.

\textbf{External dataset}

The dataset comprises 646 frontal chest X-rays, with 232 images specifically selected from the BELARUS TB portal program dataset  representing cases of TB chest X-rays. Additionally, 414 Normal chest X-ray images were collected from the RSNA dataset.

\subsection{Experiment details}
\label{app:B.2}
The experiments are conducted in a Python 3.9.13 environment on CentOS 7, utilizing PyTorch 2.0.0 with CUDA 12.0 support. The hardware setup comprises an Intel(R) Xeon(R) Platinum 8163 CPU @ 2.50GHz and 8 NVIDIA GeForce RTX 4090 GPUs, with a total memory of 512 GB.

(1)	DDPM training: The batch size is consistently set at 16. The total number of epochs is determined by a experience setting ($1e^6 * numClasses / dataSize$), considering that more classes or a smaller dataSize would necessitate additional epochs for effective training. The initial learning rate is established at $3e^{-4}$. Image size for the experiments is uniformly set at 64x64.

(2)	Classifier training: The ResNet architecture was chosen for its stable performance across various classification tasks. Other classifiers can also be utilized to validate the effectiveness of our privacy computing framework. Specifically, the chosen architecture is ResNet50, and we utilize the Adam optimization algorithm with a learning rate of $e^{-4}$.

(3)	Evaluation metrics: Accuracy is used to evaluate the classification performance. Accuracy is the most straightforward metric and represents the ratio of correctly predicted instances to the total instances.

\section{More Results}
\label{app:C}
\subsection{Tuberculosis distributions}
\label{app:C.1}
\begin{figure*}[!h]
\vskip 0.2in
\begin{center}
      \subfigure[]{
        \includegraphics[width=0.46\linewidth]{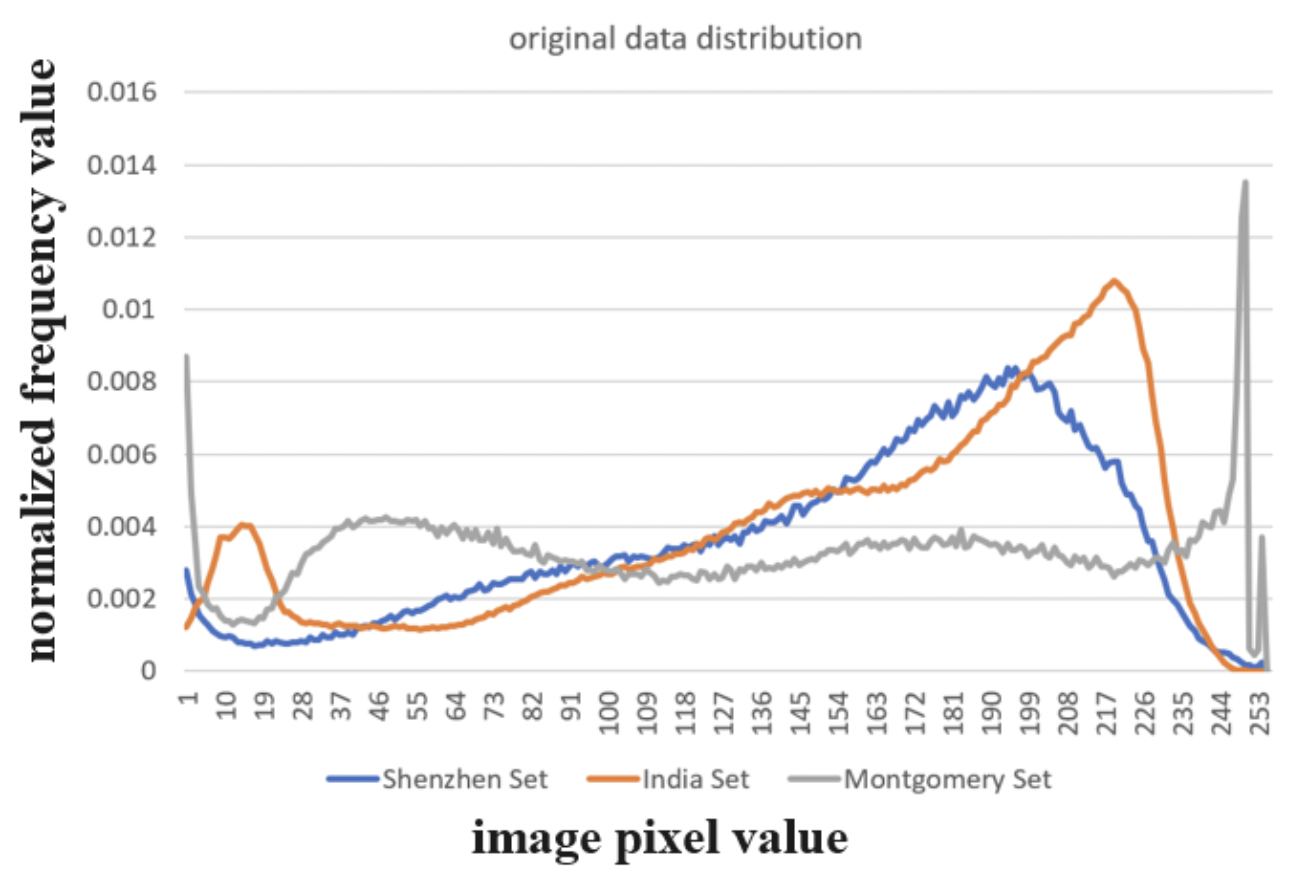}
      }
      \subfigure[]{
        \includegraphics[width=0.44\linewidth]{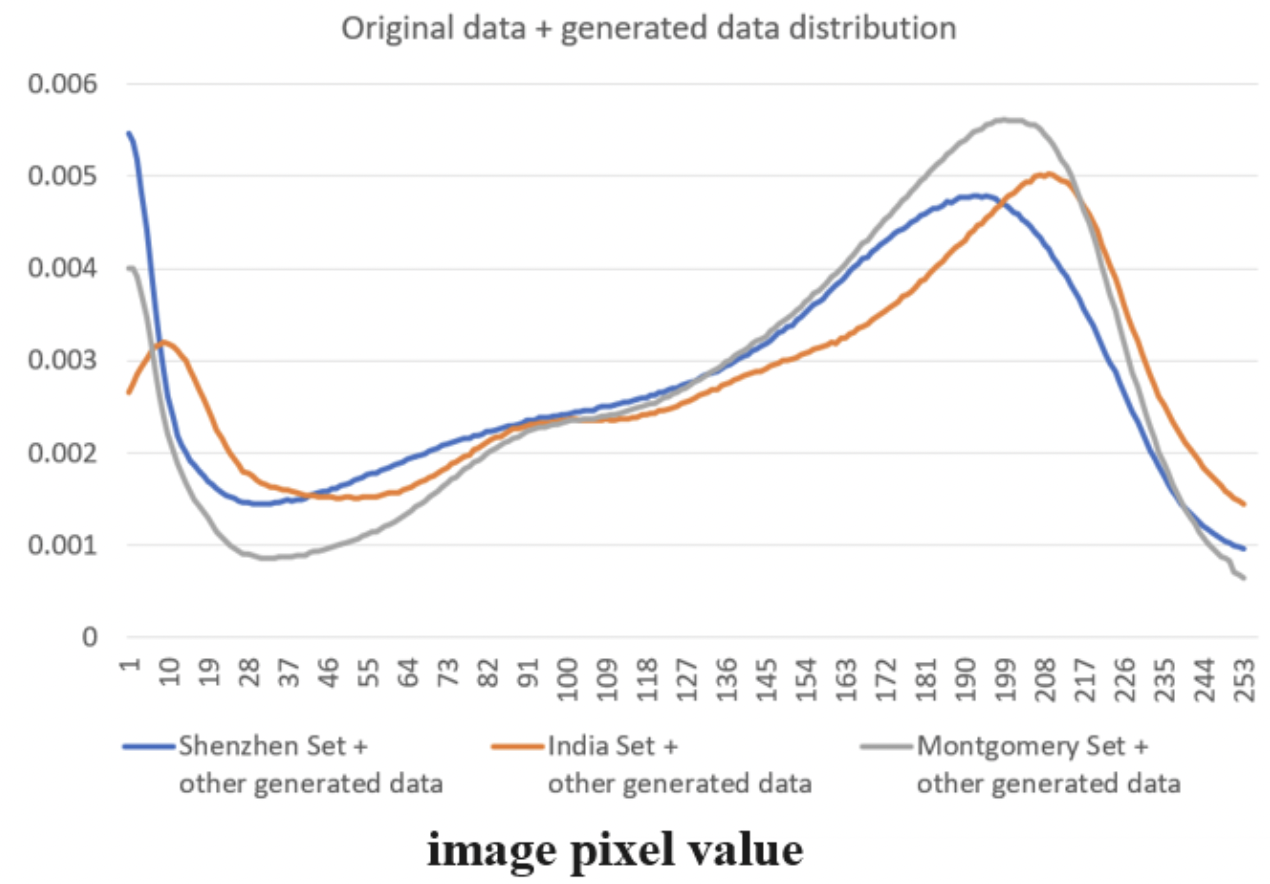}
      }
\end{center}
\caption{The image distributions of the three Tuberculosis clients.}
\label{fig-3}
\vskip -0.2in
\end{figure*}

\cref{fig-3} illustrates the distinct distributions among the three Tuberculosis clients. Handling Non-IID data across different clients poses challenges for federated learning. However, our method's generated data effectively mitigates distribution differences among clients. \cref{fig-3} (a) depicts the distribution of different clients on original data, while \cref{fig-3} (b) illustrates the distribution when original data is combined with generated data. Our method's generated data plays a crucial role in substantially reducing the distribution disparities among clients.

\subsection{Prediction Aggregation comparison}
\label{app:C.2}
For prediction aggregation, we conducted a performance comparison between the voting and averaging methods. The results indicate that there is no significant difference between the two methods, with most differences being less than 1\%, as demonstrated in \cref{table-3}. The default prediction aggregation method in this paper is averaging unless specified otherwise. 

\begin{table*}[!h]
\caption{Prediction Aggregation comparison, including voting and averaging method.}
\label{table-6}
\vskip 0.15in
\begin{center}
\begin{small}
\begin{rm}
\setlength{\tabcolsep}{3.5mm}{\begin{tabular}{lcccccccc}
\toprule
Prediction Aggregation & \multicolumn{2}{c}{IntelImage}&  \multicolumn{2}{c}{NeoJaundice} &\multicolumn{2}{c}{ChestX-Ray Pneumonia} &\multicolumn{2}{c}{Tuberculosis}\\
~ & validation & test & validation & test & validation & test & validation & test \\
\midrule
voting & 90.66 & 92.98 & \textbf{82.81} & \textbf{77.78} & 90.71 & 90.70 & \textbf{92.94} & 89.12 \\
averaging & \textbf{91.96} & \textbf{93.25} & 81.38 & 77.04 & \textbf{91.37} & 90.70 & 90.59 & \textbf{90.16} \\
\bottomrule
\end{tabular}
}
\end{rm}
\end{small}
\end{center}
\vskip -0.1in
\end{table*}

\subsection{Ablation experiment}
\label{app:C.3}
In this section, we conduct a series of ablation experiments to comprehensively analyze and measure the influence of both generated data and predictions aggregation on the performance of the classification task, as shown in \cref{table-7} and \cref{table-2}. We establish a baseline through centralized learning, exclusively training the model on the original data (A1, A2, A3). Subsequently, we enrich the training sets by incorporating generated data. Another ablation experiment explores predictions aggregation across all clients' outputs.

(1)	Centralized Learning with Original Data (A1, A2, A3)

The initial configuration serves as our baseline, representing the model trained through centralized learning exclusively on the original data. This sets the foundation for assessing the model's inherent capabilities.

(2)	Centralized Learning with Original Data + Other Clients' Generated Data (B1, B2, B3)

To explore the effects of introducing generated data from other clients, we augment the original data with data synthesized from Other Clients. This step allows us to gauge the collaborative impact of original and additional synthesized information.

Additionally, we perform (4) predictions aggregation across all original data predictions (A1 to A3), and (5) predictions aggregation across all Original Data + Other Clients' Generated Data (B1 to B3) predictions.

\begin{figure}[!h]
\vskip 0.2in
\begin{center}
\centerline{\includegraphics[width=0.8\linewidth]{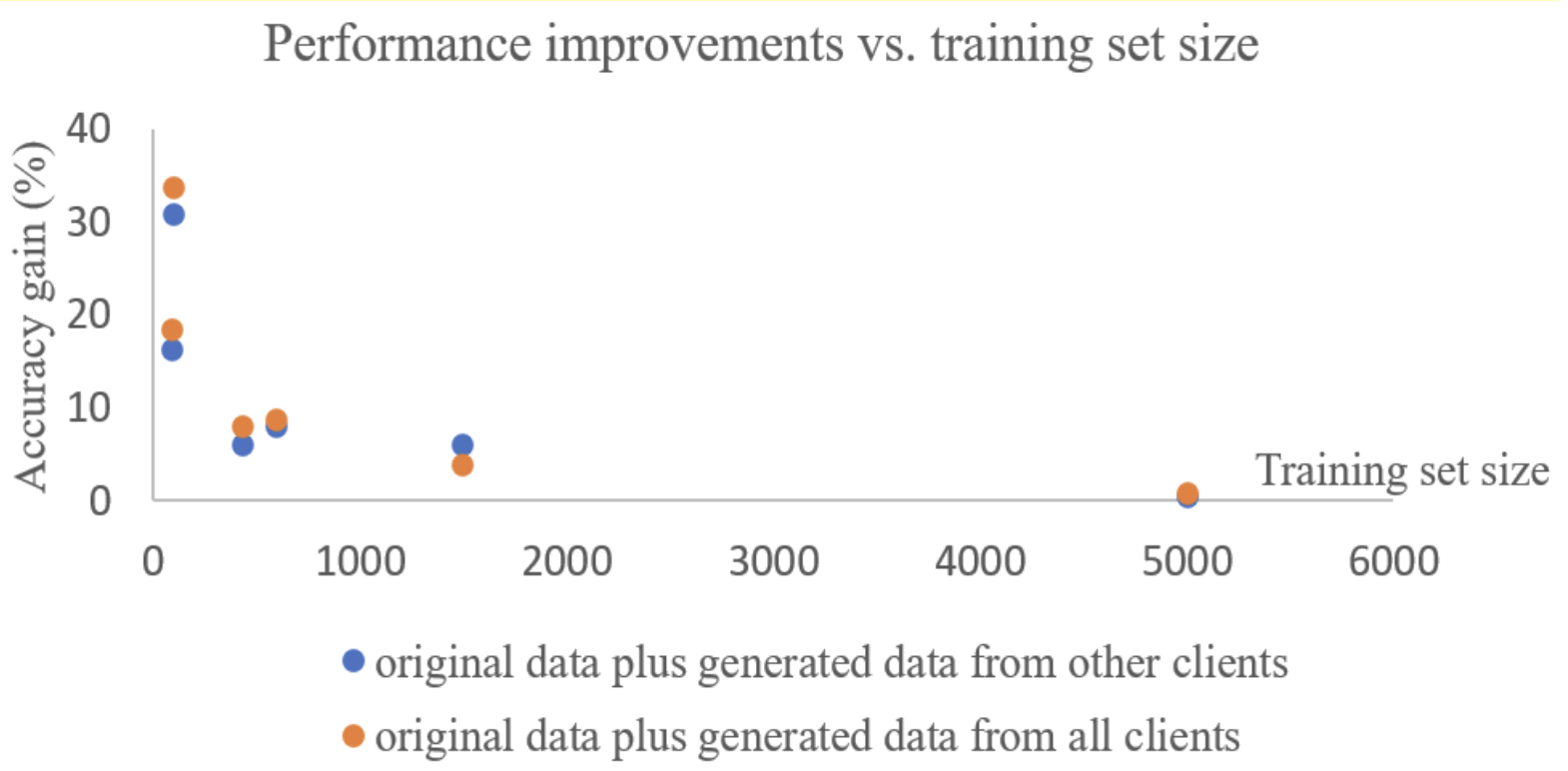}}
\caption{The relationship between performance improvements and training set size.}
\label{fig-4}
\end{center}
\vskip -0.2in
\end{figure}

\textbf{Impact of generated data}\\
In the IntelImage dataset, the addition of generated data results in only a marginal improvement ($<1\%$). Conversely, in the NeoJaundice, ChestX-Ray Pneumonia, and Tuberculosis datasets, the combination of original data with generated data from other clients (B1, B2, B3) showcases a substantial improvement, exceeding 8\% on average. This indicates that while the performance improvement in a relatively large natural dataset (IntelImage) is limited, the improvement is more significant in smaller medical datasets (NeoJaundice, ChestX-Ray Pneumonia, and Tuberculosis). As illustrated in \cref{fig-4}, there is a discernible trend wherein smaller training set sizes tend to gain more from the utilization of generated data.

\begin{figure*}[ht]
\vskip 0.2in
\begin{center}
      \begin{minipage}[b]{\linewidth}
      \subfigure[]{
        \includegraphics[width=0.46\linewidth]{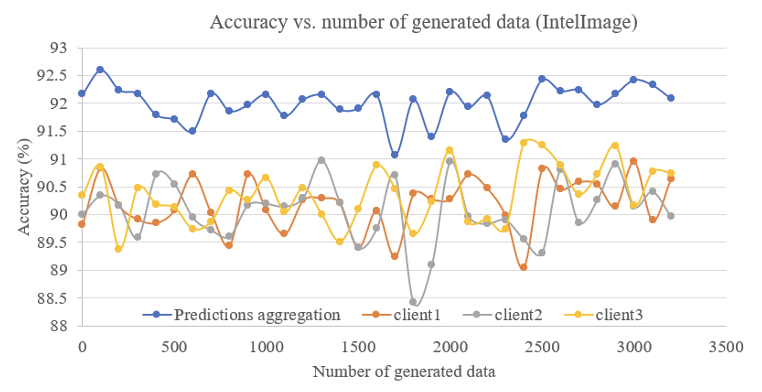}
      }
      \subfigure[]{
        \includegraphics[width=0.46\linewidth]{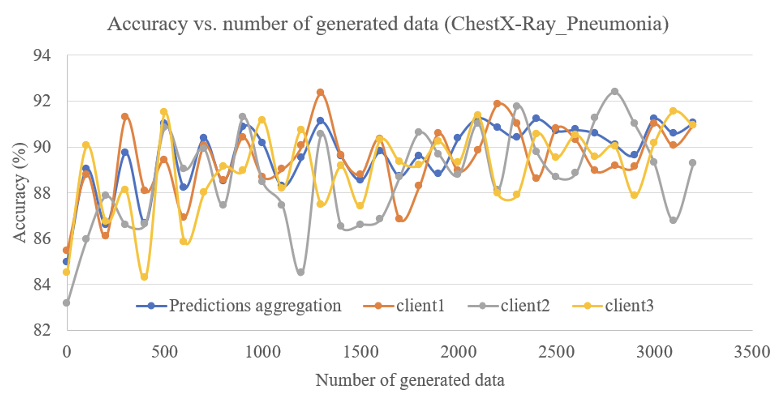}
      }
    \end{minipage}
    \begin{minipage}[b]{\linewidth}
      \subfigure[]{
        \includegraphics[width=0.46\linewidth]{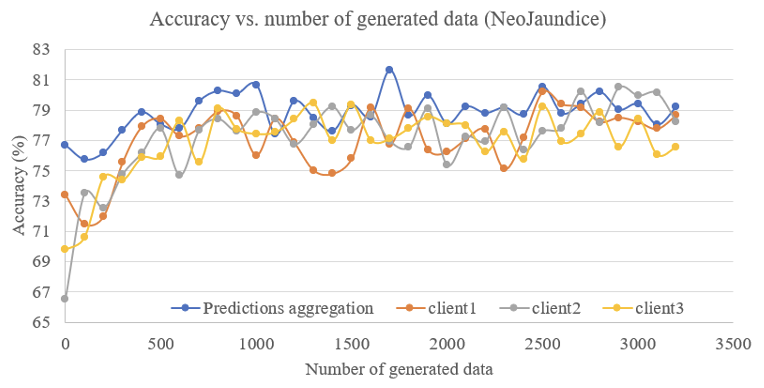}
      }
      \subfigure[]{
        \includegraphics[width=0.46\linewidth]{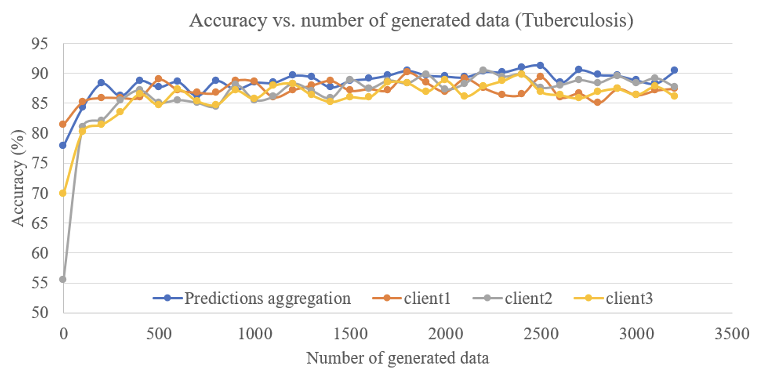}
      }
    \end{minipage}
\end{center}
\caption{The impact of the number of generated data on accuracy across the four datasets.}
\label{fig-5}
\vskip -0.2in
\end{figure*}

\textbf{Impact of Predictions aggregation}\\
From \cref{table-7}. and \cref{table-2}. it is evident that all predictions aggregations outperform the results of individual clients. The (A1, A2, A3) predictions aggregation, and (B1, B2, B3) predictions aggregation show improvements of 5.2\%, 2.0\%, respectively. This highlights the significant role predictions aggregation plays in rapidly enhancing overall performance.

\begin{table*}[!h]
\caption{The impact of the number of generated data on accuracy across the four datasets.}
\label{table-7}
\vskip 0.15in
\begin{center}
\begin{small}
\begin{rm}
\setlength{\tabcolsep}{3mm}{\begin{tabular}{llcccccc}
\toprule
Dataset & \multicolumn{2}{c}{IntelImage}&  \multicolumn{2}{c}{NeoJaundice} &\multicolumn{2}{c}{ChestX-Ray Pneumonia}\\
\hline
\addlinespace[0.5ex]
Training set & validation & test & validation & test & validation & test \\
\midrule
        Client1 (A1) & 90.03 & 89.61 & 72.49 & 74.07 & 84.07 & 86.63 \\ 
        Client1 + other generated data (B1) & 89.51 & 91.77 & 78.80 & 78.52 & 90.04 & \textbf{91.86} \\
        Client2 (A2) & 89.86 & 90.15 & 64.18 & 68.89 & 84.96 & 81.40 \\
        Client2 + other generated data (B2) & 88.84 & 91.09 & \textbf{81.66} & 74.81 & 89.60 & 88.95 \\
        Client3 (A3) & 89.86 & 90.82 & 72.21 & 67.41 & 84.73 & 84.30 \\
        Client3 + other generated data (B3) & 89.91 & 91.63 & 79.08 & 74.07 & 91.15 & 90.69 \\
\addlinespace[0.5ex]
\hline 
\addlinespace[0.5ex]
        (A1 to A3) Predictions aggregation & \textbf{92.03} & 92.31 & 76.79 & 76.30 & 85.62 & 84.30 \\
        (B1 to B3) Predictions aggregation (Ours) & 91.96 & \textbf{93.25} & 81.38 & \textbf{77.04} & \textbf{91.37} & 90.70 \\
\bottomrule
\end{tabular}
}
\end{rm}
\end{small}
\end{center}
\vskip -0.1in
\end{table*}

\textbf{Number of generated data}\\
Finally, we investigate the influence of the number of generated data on performance. The number of generated data varies from 0 to 3200, with increments of 100. Performance is quantified as the average accuracy derived from the validation and testing sets. The graphical depiction of the number of generated data's impact is presented in \cref{fig-5}.

From \cref{fig-5}, it can be observed that performance does not monotonically increase with the growth of generated data. All curves exhibit fluctuations, although the overall trend shows improvement with increasing generated data. This is because the quality of data generated by the DDPM model has a certain level of randomness, and poor-quality generated data may lead to a decline in performance. For most datasets, a performance boost is observed with the addition of just 100 generated data. For example, the IntelImage dataset shows a 0.43\% improvement, ChestX-Ray Pneumonia improves by 4.65\%, and Tuberculosis datasets exhibit a 6.45\% improvement. The only exception is NeoJaundice, which experiences a 0.94\% decrease. However, as the amount of data increases, performance begins to improve, and by the time it reaches 500, there is a 1.37\% improvement compared to the original performance.

The performance of each client fluctuates significantly with different amounts of generated data, with many experiencing fluctuations as high as 5\%. In contrast, the fluctuations in data after Predictions Aggregation are much smaller, and after reaching 2000 generated data, the fluctuations gradually decrease. This also indicates the significant role of Predictions Aggregation in maintaining stable performance.

For all datasets, when the number of generated data reaches 3200, performance tends to stabilize. While increasing the amount of generated data might potentially enhance performance further, it comes at the cost of additional time required for sample generation. Therefore, for the sake of practicality and efficiency in this experiment, the default number of generated data used is set to 3200.

\section{Generated images from the DDPM}
\label{app:D}
Training stable models with a small number of samples is a crucial challenge for DDPM, especially when starting from scratch without pre-trained models, we generate stable and high-quality models for different datasets. While diffusion-generated samples exhibit high quality, training a model solely on generated data to achieve performance similar to models trained on original data remains challenging. Both natural images and medical images are selected for our experiments.
\begin{figure*}[!h]
\vskip 0.2in
\begin{center}
\centerline{\includegraphics[width=0.8\textwidth]{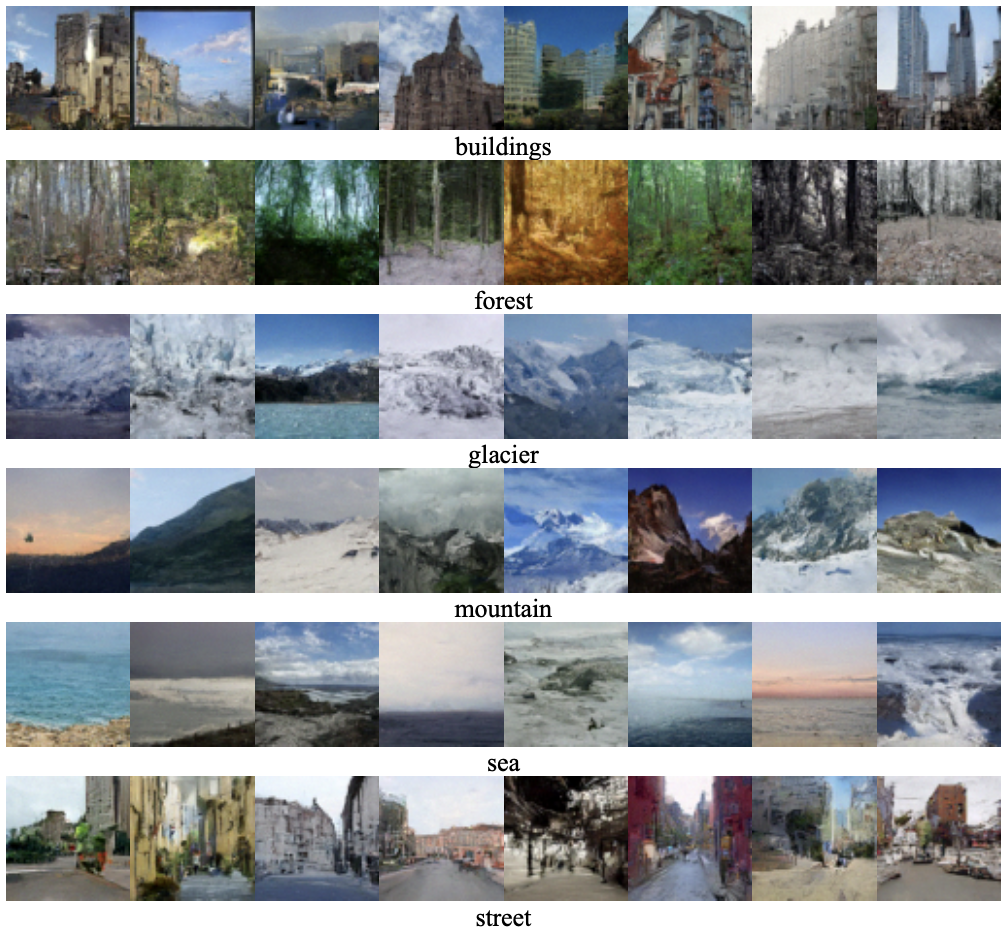}}
\caption{IntelImage}
\label{app-1}
\end{center}
\vskip -0.2in
\end{figure*}

\begin{figure*}[!h]
\vskip 0.2in
\begin{center}
\centerline{\includegraphics[width=0.8\textwidth]{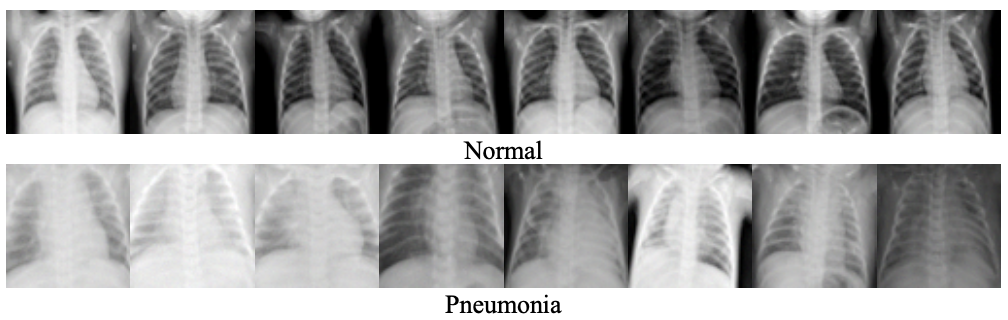}}
\caption{ChestX-Ray Pneumonia}
\label{app-2}
\end{center}
\vskip -0.2in
\end{figure*}

\begin{figure*}[t]
\vskip 0.2in
\begin{center}
\centerline{\includegraphics[width=0.8\textwidth]{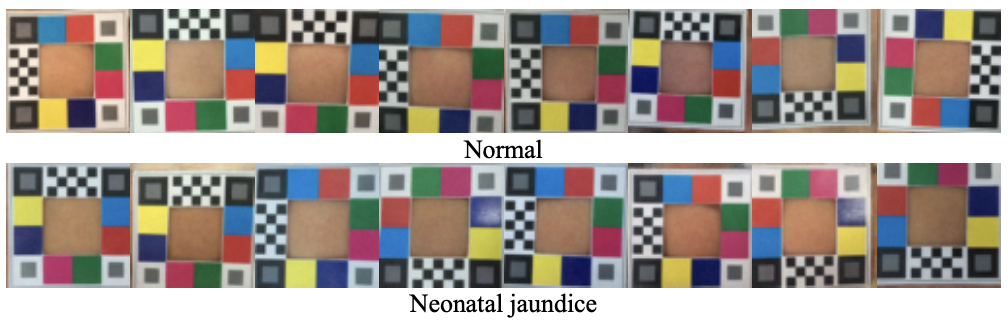}}
\caption{NeoJaundice}
\label{app-3}
\end{center}
\vskip -0.2in
\end{figure*}

\begin{figure*}[t]
\vskip 0.2in
\begin{center}
\centerline{\includegraphics[width=0.8\textwidth]{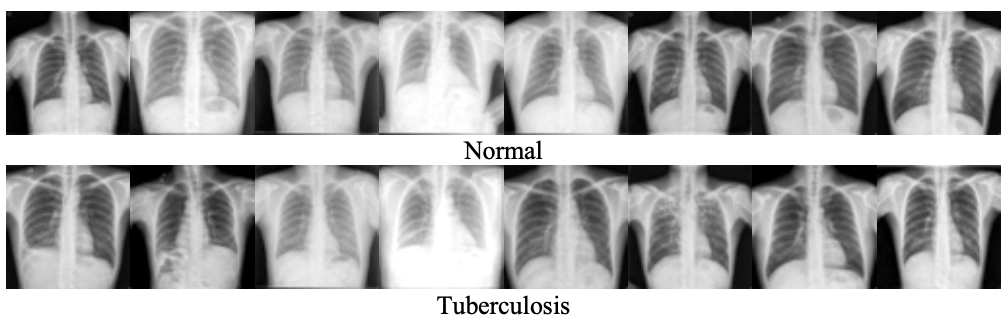}}
\caption{Tuberculosis}
\label{app-4}
\end{center}
\vskip -0.2in
\end{figure*}


\end{document}